\documentclass[showpacs,preprintnumbers,amsmath,amssymb,prl]{revtex4}    %  twocolumn,
\usepackage{graphicx}% Include figure files
\usepackage{dcolumn}% Align table columns on decimal point
\usepackage{bm}% bold math
\usepackage{color}

\newcommand{\bea}{\begin{eqnarray}}
\newcommand{\eea}{\end{eqnarray}}
\newcommand{\be}{\begin{equation}}
\newcommand{\ee}{\end{equation}}

\begin{document}
\title{Enumeration of RNA complexes via random matrix theory}

%\author{Piotr Su{\l}kowski}

\author{J{\o}rgen E. Andersen$^{1}$, Leonid O.\ Chekhov$^{3,4}$,  R.\ C.\ Penner$^{1,2}$,
Christian M. Reidys$^{5}$, Piotr Su{\l}kowski$^{2,6,7}$ \\ 
}

\affiliation{
$^1$ Center for Quantum Geometry of Moduli Spaces, Aarhus University, DK-8000 {\AA}rhus C, Denmark \\
$^2$ Division of Physics, Mathematics and Astronomy, \\ California Institute of Technology, Pasadena, California, 91125 USA \\
$^3$ Department of Theoretical Physics, Steklov Mathematical Institute, Moscow, 119991 Russia \\
$^4$ School of Mathematics, Loughborough University, Leicestershire, UK \\
$^5$ Department of Mathematics and Computer Science, University of Southern Denmark, DK-5230 Odense M, Denmark \\
$^6$ Institute for Theoretical Physics, University of Amsterdam, \\ Science Park 904, 1090 GL, Amsterdam, The Netherlands \\
$^7$ Faculty of Physics, University of Warsaw, \\ul. Ho{\.z}a 69, 00-681 Warsaw, Poland \\
}

%\date{\today}

%\begin{abstract}

%\end{abstract}


\begin{flushright}
CALT-68-2896
\end{flushright}

\maketitle

%\section{Introduction}

{\bf
We review a derivation of the numbers of RNA complexes of an arbitrary topology. These numbers are encoded in the free energy of the hermitian matrix model with potential $V(x)=x^2/2-stx/(1-tx)$, where $s$ and $t$ are respective generating parameters for the number of RNA molecules and hydrogen bonds in a given complex. The free energies of this matrix model are computed using the so-called topological recursion, which is a powerful new formalism arising from random matrix theory. These numbers of RNA complexes also have profound meaning in mathematics: they provide the number of chord diagrams of fixed genus with specified numbers of backbones and chords as well as the number of cells in Riemann's moduli spaces for bordered surfaces of fixed topological type.

\medskip

%We introduce and study the Hermitian matrix model with potential $V_{s,t}(x)=x^2/2-stx/(1-tx)$, which enumerates the number of linear chord diagrams with no isolated vertices of fixed genus with specified numbers of backbones generated by $s$ and chords generated by $t$. For the one-cut solution, the partition function, correlators and free energies are convergent for small $t$ and all $s$ as a perturbation of the Gaussian potential, which arises for $st=0$.  This perturbation is computed using the formalism of the topological recursion.  The corresponding enumeration of chord diagrams gives at once the number of RNA complexes of a given topology as well as the number of cells in Riemann's moduli spaces for bordered surfaces. The free energies are computed here in principle for all genera and explicitly in genus less than four.
}

\bigskip

%******************************************
%******************************************

\begin{figure}[htb]
\begin{center}
\includegraphics[width=0.5\textwidth]{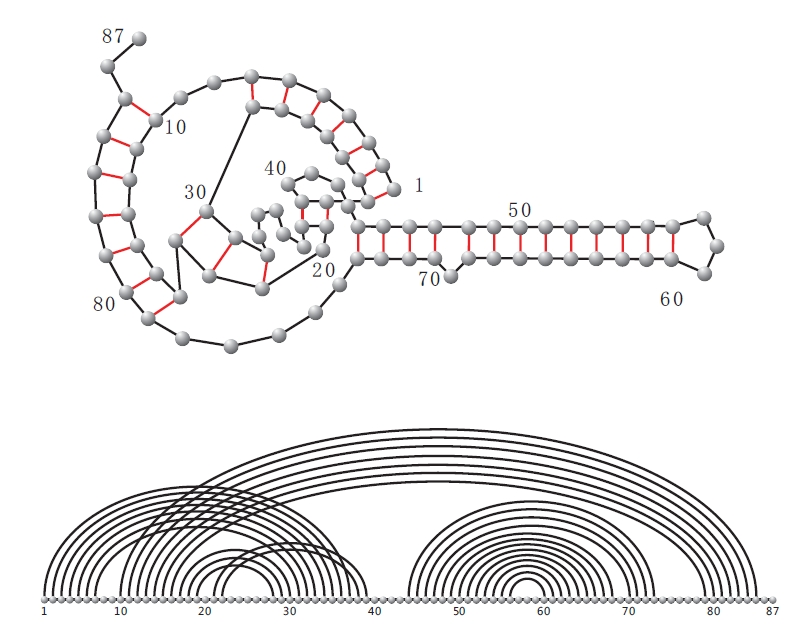}
\caption{RNA complex (above) can be uniquely represented as a chord diagram (below). Each RNA chain (just one in the example in the figure) corresponds to a single backbone (a horizonatal interval in the chord diagram), and each hydrogen bond is represented by a chord (a semi-circle with ends attached to a backbone).
} \label{fig-RNAchord}
\end{center}
\end{figure}

In this paper, we review a solution to the following problem studied and solved in \cite{RNAchords}. Let us consider a complex consisting of an arbitrary number of RNA chains with various nucleotides connected by Watson-Crick bonds both intra- and inter-chain. We seek the number of such topologically inequivalent complexes consisting of a given number of RNA chains and bonds. To explain the topology, it is useful to rephrase this problem in terms of the so-called chord diagrams. Chord diagrams are comprised of a number of so-called backbones, which are represented by disjoint, oriented and labeled intervals lying along a fixed line, that are connected by so-called chords taken as
semi-circles lying above this fixed line. Each backbone is identified with the sugar-phosphate backbone (hence the terminology) of a single RNA molecule oriented from its 5' to 3' end. 
Chords correspond to Watson-Crick basepairs, where we add an associated chord taking care its endpoints in each backbone occur in the correct order corresponding to the primary structure, i.e., the word in the four-letter alphabet of nucleic acids that determines the RNA molecule.  Endpoints of chords lie at distinct interior points of the backbones. 
In this way, a complex of interacting RNA molecules determines a chord diagram. We assume that nucleotides not participating in basepairs play no role in this model, i.e., there are no isolated vertices, and that chord diagrams are connected, i.e., the corresponding RNA complexes are also connected. Now, distinct isomorphism classes of chord diagrams correspond to the topologically inequivalent configurations we wish to count. An example of an RNA configuration (consisting of a single RNA chain) and the corresponding chord diagram (on one backbone), are shown in Fig. \ref{fig-RNAchord} (this figure is borrowed from \cite{gfold}). We also stress that the results presented in what follows can be applied to various complexes of (bio)polymers, not necessarily RNA complexes. However, for definiteness, we are going to discuss these results just in the RNA context.

\begin{figure}[htb]
\begin{center}
\includegraphics[width=0.7\textwidth]{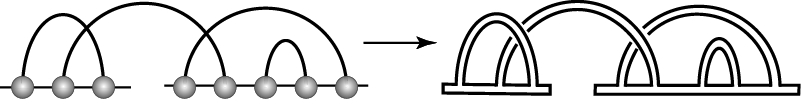}
\caption{Each RNA complex uniquely determines a chord diagram consisting of $b$ backbones and $n$ chords (left). By thickening backbones and chords we obtain a  surface (right) with $r$ boundary components. The genus $g$ of this surface  provides an important characteristic of the corresponding chord diagram, and it can be determined from the Euler relation $b-n=2-2g-r$. Our task is to determine the numbers $c_{g,b}(n)$ of topologically inequivalent chord diagrams on $b$ backbones, with $n$ chords, and characterized by genus $g$.
} \label{fig-surface}
\end{center}
\end{figure}

It turns out that the counting of RNA complexes, or chord diagrams, can be refined by introducing an additional parameter, which naturally characterizes the topology of these structures. This parameter arises as the genus $g\geq 0$ of an auxiliary topological surface with boundary which can be associated to a given chord diagram by thickening its backbones and chords, as shown in Fig. \ref{fig-surface}. This surface has some number $r\geq 1$ of boundary components. If we denote the number of backbones and chords respectively by $b$ and $n$, then the Euler characteristic of these auxiliary surface can be expressed as $b-n=2-2g-r$; this relation can be used in particular to determine the genus $g$. For example in fig. \ref{fig-surface}, we have $b = 2$ backbones, $n = 4$ chords, and $r = 4$ boundary components, which implies that the corresponding surface has genus $g = 0$.  Let us remark
parenthetically that \cite{gfold} provides a context-free grammar and polynomial algorithm for computing minimum-free energy configurations of a single RNA molecule while allowing for certain
pseudo-knot patterns of arbitrarily high genus that arise by suitably iterating genus one; since the $c_{g,b}(n)$ grow rapidly in $g$ or $b$ for $n$  in the appropriate range, it is unlikely that a similar higher-genus approach is reasonable short of a full-blown and as-yet-unknown field theory for RNA.

To sum up, a given RNA complex or chord diagram is characterized by its number of backbones $b$, its number of chords $n$, and its genus $g$. Let us denote the number of topologically inequivalent such diagrams by $c_{g,b}(n)$. These are the numbers we wish to determine. In what follows, we explain how to determine the following generating functions of these numbers
\be\label{C-intro}
C_{g,b}(z)=\sum _{n\geq 0}c_{g,b}(n)~z^n, ~{\rm for}~g\geq 0,
\ee
using random matrix theory, and in particular, the so-called topological recursion. Before we present how these generating functions can be found, let us provide some examples. In the special case of one backbone and genus $g=0$, it is not hard to see that $c_{0,1}(n)$ are Catalan numbers with the generating function
\be
C_{0,1}(z) = {{1-\sqrt{1-4z}}\over{2z}} = 1 + z + 2z^2 + 5z^3 + 14z^4 + \ldots  \label{C0-Catalan}
\ee
so that $c_{0,1}(0)=c_{0,1}(1)=1$, $c_{0,1}(2)=2$, $c_{0,1}(3)=5$, and so on. More generally, configurations which involve only one backbone can be enumerated by the simplest matrix model with the quadratic (Gaussian) potential \cite{OrlandZee02}. However this Gaussian model cannot describe configurations involving many backbones, and constructing a model which works for many backbones is an important motivation for our work; we describe this new model below. Using this new model, we find for example that generating functions of diagrams in genus zero and one on four backbones with an arbitrary number of chords take the form
\be
\aligned
& C_{0,4} (z)= \frac{24 z^3 (3 + 18 z + 8 z^2)}{(1 - 4 z)^{7}} =  72 z^3 + 2448 z^4 + \ldots  \\
& C_{1,4} (z)= \frac{24 z^5 (715 + 7551z + 12456 z^2 + 2096 z^3) }{(1 - 4 z)^{10}} = 17160 z^5 + \dots  \label{C0414intro}
\endaligned
\ee
This means that $c_{0,4}(3)=72$, $c_{0,4}(4)=2448$, for example. While computation of these numbers by explicit enumeration of chord diagrams is possible for relatively low and fixed $g,b$ and $n$, as illustrated, e.g., in the appendix in \cite{RNAchords}, it quickly 
becomes involved. On the other hand, the random matrix theory allows us to determine generating functions $C_{g,b}(z)$ in an algorithmic way, in principle for all $g$ and $b$. As yet another example, for 4 and 5 backbones, in genus 2 and 3, we find the following generating functions
\bea
C_{2,4}(z) & \, = \,& \frac{144 z^7}{(1 - 4 z)^{13}} (38675 + 620648 z + 2087808 z^2 \nonumber \\
    && + 1569328 z^3 +  134208 z^4)  , \nonumber  \\
C_{3,4}(z) & \, = \,&  \frac{48z^9}{ (1 - 4 z)^{16}} (53416125 + 1194366915 z + 6557325096 z^2 \nonumber \\
    && + 10738411392 z^3 + 4580024832 z^4 + 236239616 z^5),   \nonumber\\
C_{2,5}(z) & \, = \,& \frac{144 z^8}{(1 - 4 z)^{{31}\over 2}} (2543625 + 62424520 z + 375044396 z^2 \label{Cexamples} \\
    && + 671666053 z^3 + 314761848 z^4 + 18335696 z^5)  , \nonumber  \\
C_{3,5}(z) & \, = \,&  \frac{720z^{10}}{ (1 - 4 z)^{{37}\over 2}} (360380790 + 11275076865 z + 95744892585 z^2  \nonumber \\
    && + 282797424880 z^3 + 291167707410 z^4 + 85497242928 z^5 + 3218434848 z^6).   \nonumber
\eea

%********************************************************************
%********************************************************************

We can explain now how the above generating functions can be determined from the random matrix theory. As shown in \cite{RNAchords}, all chord diagrams can be enumerated by the following matrix model, i.e., an integral over hermitian matrices $H$ of size $N$,
\be
Z=\int DH~e^{-N{\rm tr}V(H)} = {\rm exp}~\biggl (-N^2s+{\sum_{g=0}^{\infty} N^{2-2g} F_g}\biggr ),   \label{Zintro}
\ee
where the so-called matrix model potential is given by
\be
V(x)={{~x^2}\over 2}-{{stx}\over{1-tx}}.   \label{Vmatrix-intro}
\ee
These expressions should be understood as follows. The integral in (\ref{Zintro}) with the potential given in (\ref{Vmatrix-intro}) arises from the combinatorial analysis of chord diagrams, and its form was found in \cite{RNAchords} using well-known techniques in random matrix models, analogous, e.g., to those used in \cite{Penner88}. In the limit of large matrix size $N\to \infty$, the free energy of this model $F=\log Z$ has the genus expansion in $N$ given in the exponent on the right side of (\ref{Zintro}) (the extra summand $-N^2s$ in the exponential in (\ref{Zintro}) is purely for convenience). In general, finding the free energy of a matrix model of the above form (with arbitrary potential $V(x)$) is a very difficult task. On the other hand, such free energies $F_g$, for the potential given in (\ref{Vmatrix-intro}), are essentially the objects we are after. Namely, from the combinatorial description discussed in \cite{RNAchords}, it follows that the free energies $F_g$ encode the generating functions (\ref{C-intro})
\be
\label{eq:freeg}
F_g(s,t)= const + \sum_{b\geq 1} {{s^b}\over{b!}}~C_{g,b}(t^2),~{\rm for}~g\geq 0,
\ee
where the constant terms reproduce the free energies $ \frac{B_{2g}}{2g(2g-2)}$ of the Gaussian matrix model (characterized by the quadratic potential $V(x)=\frac{1}{2}x^2$), where $B_{2g}$ denote Bernoulli numbers. The extra factor $b!$ arises because $c_{g,b}(n)$ counts chord diagrams with labeled backbones as opposed to unlabeled ones as arise in the matrix model description (and where a permutation of backbones must nevertheless preserve backbone orientations). We also see that the genus expansion into free energies $F_g$ agrees with the genus $g$ of the auxiliary surfaces described earlier.

The problem of enumerating of RNA complexes therefore reduces to the problem of performing the matrix integral and determining free energies $F_g$ in (\ref{Zintro}). As we already mentioned, in general this is a very difficult problem; however, recently a beautiful algorithmic solution to this problem has been given. To find free energies one should solve the so-called loop equations of the matrix model, which are equations (known as Ward identities in quantum field theory) satisfied by certain multi-linear correlators $W^{(g)}_n(p_1,\ldots,p_n)$ in this model. The leading order equation among those identities specifies a so-called spectral curve, i.e., an algebraic curve which characterizes distribution of eigenvalues in the matrix model in the $N\to\infty$ limit. It also turns out that all correlators $W^{(g)}_n(p_1,\ldots,p_n)$ and loop equations they satisfy can be encoded entirely in terms of this spectral curve. These loop equations can be solved in a recursive way \cite{ChekhovEynard05,ChEO06}, and in this manner, free energies $F_g$ (for $g\geq 2$) are completely determined by correlators $W^{(g)}_1(p)$. This entire procedure requires just the knowledge of the spectral curve, which can be regarded as the initial condition of the recursion (and a universal form of the solution to loop equations) and no other details of a matrix model from which this curve was derived. An important achievement of Eynard and Orantin \cite{EO07} was to realize that one can use the recursive solution of loop equations to assign correlators $W^{(g)}_n(p_1,\ldots,p_n)$ and $F_g$ to an arbitrary algebraic curve, not necessarily of matrix model origin. On the other hand, it is guaranteed that $F_g$ computed for the spectral curve of a matrix model reproduce the free energies.

In order to solve the matrix model (\ref{Zintro}) with the potential (\ref{Vmatrix-intro}), we can therefore use the formalism of this topological recursion. This has indeed been done in \cite{RNAchords}, and the main steps of this solution are as follows. First, we need to determine the spectral curve of the model (\ref{Zintro}). This can be done by the analysis of a distribution of eigenvalues in the large $N$ limit. Because the potential (\ref{Vmatrix-intro}) is a deformation of the quadratic function, it has a single minimum, and in the equilibrium configuration eigenvalues spread around this minimum. For large $N$ the eigenvalues are distributed along an interval with end-points $a$ and $b$, which defines a cut in a certain auxiliary complex plane. Such a one-cut solution defines the corresponding spectral curve which has genus zero, and it turns out to be given by the following algebraic equation for two complex variables $x$ and $y$
\be
4 y^2 (tx-1)^4 = (x-a)(x-b) \Big( \big(tx - 1 + \frac{(a+b)t}{4}\big)^2 + \gamma\Big)^2,      \label{spectral-xy}
\ee
where
\be
\gamma  = - \frac{(at + bt) \big( (at)^2 + (bt)^2 + 14(at+bt-abt^2) -16 \big)}{16 (at + bt - 2)}.
\ee
While the end-points of the cut $a$ and $b$ cannot be given in a closed form, it can be found that they are determined by the following system of equations
\be
\left\{\begin{array}{l} 0 = a + b + \frac{st(at+bt-2)}{\big((at-1)(bt-1) \big)^{3/2}},      \\
16 = (a-b)^2 + \frac{4s \big((2-\frac{(a+b)t}{2})(at+bt-2) + 2abt^2 - 3t(a+b) + 4  \big)}{\big((at-1)(bt-1) \big)^{3/2}}.   \end{array} \right.   \label{ab-equations}
\ee
From the knowledge of the curve (\ref{spectral-xy}) and the formalism of the topological recursion we can now determine $F_g$ for $g\geq 2$ ($F_0$ and $F_1$ must be determined separately, independently of the recursion). In particular, we find the following exact result for the free energy at genus 2:
\be
\aligned
F_2 & = -\frac{t^4 (1-\sigma)^2}{240 \delta^4 (1 - \delta - 4 \sigma + 3 \sigma^2)^5 (1 + \delta - 4 \sigma + 3 \sigma^2)^5} \times \nonumber \\
& \times \Big(160 \delta^4 (1 - 3 \sigma)^4 (1 - \sigma)^6 
- 80 \delta^2 (1 - 3 \sigma)^6 (1 - \sigma)^8   \nonumber\\
& + 16 (1 - 3 \sigma)^8 (1 - \sigma)^{10} + \delta^{10} (-16 + 219 \sigma - 462 \sigma^2 + 252 \sigma^3)   \nonumber \\
& + 10 \delta^6 (1 - 3 \sigma)^2 (1 - \sigma)^4 (-16 - 126 \sigma - 423 \sigma^2 + 2286 \sigma^3 - 
     2862 \sigma^4 + 1134 \sigma^5) \nonumber \\
& + 5 \delta^8 (1 - \sigma)^2 (16 + 189 \sigma - 2970 \sigma^2 + 9549 \sigma^3 - 11286 \sigma^4 + 4536 \sigma^5)\Big)  \nonumber
\endaligned
\ee
where
\be
\sigma = \frac{(a+b)t}{2},\qquad\quad \delta = \frac{(a-b)t}{2}.   \label{abSD}
\ee
We also obtain an exact result for the free energy $F_3$ which is yet more complicated, with its precise form given in \cite{RNAchords}. 
Expanding these results in the form given in (\ref{eq:freeg}) and using the perturbative expansion of $a$ and $b$ in $s$ which follows from (\ref{ab-equations}), we can determine appropriate generating functions $C_{g,b}(z)$, such as those given in (\ref{Cexamples}). This procedure can be continued in an algorithmic manner, and with sufficient computational power, one can determine exact forms of $F_g$ for any $g$, and so the corresponding $C_{g,b}(z)$, and finally all $c_{g,b}(n)$.

To sum up, we have shown how random matrix theory and the topological recursion can be used to enumerate topologically inequivalent RNA complexes. This result opens many other perspectives and research possibilities. On one hand, one can consider asymptotics of the numbers $c_{g,b}(n)$ which we have found, and analyze their statistical properties. One can also compare these theoretical predictions with the structure of RNA configurations observed in Nature. In our discussion, a relation to matrix models arises naturally if we translate RNA configurations into the form of chord diagrams, and therefore $c_{g,b}(n)$ at the same time count the number of such diagrams. Such chord diagrams arise in many other problems in mathematics and physics (e.g., in knot theory or algebraic geometry) and our results should shed new light on those other fields as well. 
%Moreover, our results have yet another interpretations and applications. 
In particular, a simple transform of the numbers $c_{g,b}(n)$ count a sub-class of chord diagrams called ``shapes'', which give the number of cells in the ideal cell decomposition \cite{Pennerbook12} of Riemann's moduli space for a surface of genus $g\geq 0$ with $b\geq 1$ boundary components provided $2g+b>2$. The computation reported here is therefore at once of significance in computational biology and in geometry, and represents a remarkable confluence of biology, mathematics and physics.

%********************************************************************
%********************************************************************

\bigskip 

This work has been supported by the Danish National Research Foundation center of excellence grant ``Center for Quantum Geometry of Moduli Spaces'',
the Marie-Curie IOF Fellowship and the Foundation for Polish Science. P.S. thanks the Isaac Newton Institute for Mathematical Sciences, Cambridge, for hospitality.

%********************************************************************
%********************************************************************

\end{document}